\documentclass{article}
\usepackage{amsmath,amssymb}
\usepackage{graphicx}
\usepackage{quantikz}
\usepackage{amsthm}
\usepackage[numbers,sort&compress]{natbib}
\usepackage{hyperref}
\usepackage{subcaption}
\usepackage{authblk}

\title{Quantum Random Feature Method for Solving Partial Differential Equations}

\author[3]{Junpeng Hu\thanks{hujunpeng@unitarylab.com.cn}}
\author[1,2]{Shi Jin\thanks{shijin-m@sjtu.edu.cn}}
\author[1,2,4]{Nana Liu\thanks{nana.liu@quantumlah.org}}
\author[1,2]{Lei Zhang\thanks{lzhang2012@sjtu.edu.cn}}
\affil[1]{School of Mathematical Sciences, Shanghai Jiao Tong University, Shanghai, 200240, P. R. China}
\affil[2]{Institute of Natural Sciences, MOE-LSC, Shanghai Jiao Tong University, Shanghai, 200240, P. R. China}
\affil[3]{UnitaryLab Quantum Technology Co., Ltd., Shanghai, P.R. China}
\affil[4]{Global College, Shanghai Jiao Tong University, Shanghai, P.R. China}

\DeclareMathOperator{\diag}{diag}
\DeclareMathOperator{\poly}{poly}

\newtheorem{mythm}{Theorem}
\newtheorem{mylemma}[mythm]{Lemma}
\newtheorem{myassump}[mythm]{Assumption}
\newtheorem{myexamp}{Example}

\begin{document}

\maketitle

\begin{abstract}
Quantum computing holds significant promise for scientific computing due to its potential for polynomial to even exponential speedups over classical methods, which are often hindered by the curse of dimensionality. While neural networks present a mesh-free alternative to solve partial differential equations (PDEs), their accuracy is difficult to achieve since one needs to solve a high-dimensional non-convex optimization problem using the stochastic gradient descent method and its variants, the convergence of which is difficult to prove and cannot be guaranteed. The classical random feature method (RFM) effectively merges advantages from both classical numerical analysis and neural network based techniques, achieving spectral accuracy and a natural adaptability to complex geometries. In this work, we introduce a quantum random feature method (QRFM) that leverages quantum computing to accelerate the classical RFM framework. Our method constructs PDE solutions using quantum-generated random features and enforces the governing equations via a collocation approach. A complexity analysis demonstrates that this hybrid quantum-classical algorithm can achieve a quadratic speedup over the classical RFM.
\end{abstract}

\section{Introduction}

In recent years, quantum computing has attracted substantial attention, primarily due to its potential to deliver polynomial even exponential-level acceleration relative to classical computational approaches. Within the field of scientific computing, classical numerical methods encounter inherent challenges when addressing high-dimensional problems, largely attributed to the “curse of dimensionality''. In contrast, quantum computing has demonstrated significant potential for solving partial differential equations (PDEs) in extremely high-dimensional spaces \cite{berry2014high, cao2013quantum, childs2021high, liu2021efficient, jin2022time, jin2022quantum, jin2022timenonlinear, an2023linear, an2023quantum}. For example, the Schr\"odingerisation method, first proposed by Jin et al. in 2022 \cite{jin2022quantum,jin2022quantumdetail}, provides a simple and general framework for transforming any linear dynamical system, encompassing both ordinary differential equations (ODEs) and PDEs, into Schr\"odinger-type PDEs with unitary evolution, in one  higher-dimensional space. This method has since been successfully applied to a broad spectrum of equations \cite{jin2023analog,jin2024inhomogeneous,cao2023nonautonomous,hu2024circuit,hu2024multiscale}, and can achievement optimal complexity in matrix queries with sufficiently smooth initialization \cite{jin-optimal}.

In parallel, the remarkable advancements of neural network models across a diverse range of artificial intelligence tasks have fueled a growing interest in their application to partial differential equation (PDE) solving \cite{han2017deep,weinan2018ritz,weinan2021algorithms,lee1990neural,sirignano2018dgm,raissi2019physics,zang2020weak}; notably, several neural networks exhibit the unique capability to parameterize PDE solution operators \cite{khoo2021solving,li2020fourier,lu2021learning}. Additionally, the random feature methods (RFM) \cite{rahimi2007random,liu2021random} have seen promising applications in solving PDEs \cite{chen2022bridging, chen2024optimization}, by integrating the advantages of classical algorithms and neural network approaches. While neural networks present a mesh-free alternative to solve partial differential equations (PDEs), their accuracy is difficult to achieve since one needs to solve a high-dimensional non-convex optimization problem using the stochastic gradient descent method and its variants, the convergence of which is difficult to prove and cannot be guaranteed. Thanks to its linearity, the RFM achieves spectral accuracy and employs a mesh-free framework, thereby facilitating effective applications in domains characterized by complex geometries. Additionally, the convergence properties of such linearized shallow networks have been the subject of prior research \cite{liu2025integral}. This idea bears close resemblance to extreme learning machines (ELMs) \cite{huang2004extreme,huang2011extreme} and reservoir computers (RCs) \cite{angelatos2021reservoir,lukovsevivcius2009reservoir}, which are computational paradigms that leverage fixed, nonlinear dynamics to efficiently extract information from a given dataset. Recently, quantum counterparts to ELMs and RCs (hereafter referred to as QELMs and QRCs, respectively) have garnered significant attention, owing to their potential for processing quantum information \cite{innocenti2023potential,xiong2025fundamental,monaco2024quantum,ghosh2019quantum,ghosh2021realising,kutvonen2020optimizing}.

The primary objective of this paper is to introduce a quantum random feature method (QRFM) to solve PDEs. This proposed method preserves the core principles of classical RFM while harnessing quantum computing capabilities to reduce computational complexity. Fundamentally, the core concepts of RFM involve employing random feature functions to represent approximate solutions, and adopting the collocation method to address PDEs as well as boundary conditions in the least-squares sense, thus avoiding the need for parameter training, a requirement of many other quantum neural networks \cite{ivashkov2024qkan,guo2024quantum,xiao2024physics,cao2025quantum,markidis2022physics}. Consequently, the original problem is transformed into a {\it linear} problem with respect to the coefficients of the random feature basis. Unlike nonlinear neural networks widely used in machine learning and deep neutal networks,  the linearity of RFM helps to develop quantum RFMs since quantum computers, designed by quantum mechanics principle, is mostly suitable to solve linear problems. Complexity analysis demonstrates that the proposed method achieves at least a quadratic advantage over classical RFM. In contrast to quantum extreme learning machines (QELMs) and quantum reservoir computers (QRCs) where the original problem is converted into a linear regression between measurements and targets, our proposed method centers on the quantum state and outputs a numerical solution state encoded via amplitude.

This paper is organized as follows. The basic concepts of the random feature method (RFM) are introduced in Section \ref{sec:RFM}, with the Helmholtz equation serving as an illustrative example. Section \ref{sec:quantumRFM} delineates the quantum routine derived from oracles and presents a detailed complexity analysis. The results of numerical experiments are reported in Section \ref{sec:experiment}. In Section \ref{sec:kernel}, we further highlight our method's dual interpretation via kernel methods. Finally, a general discussion and concluding remarks are provided in Section \ref{sec:discussion}.

\section{The Random Feature Method}\label{sec:RFM}

In this section, we introduce the random feature method (RFM) as proposed in \cite{chen2022bridging}. A random feature function refers to a function equipped with randomly generated feature vectors. Within the framework of machine learning methods, this concept corresponds to the step of randomly initializing network weights. 

From the perspective of numerical mathematics, the RFM method approximates the solution as a linear combination of \( M \) network basis functions \(\{\phi_m\}\) defined on the domain \(\Omega\):
\begin{equation}\label{eqn:rfm}
u_M(\boldsymbol{x}) = \sum_{m=0}^{M-1} v_m \phi_m(\boldsymbol{x}),
\end{equation}
where the basis function $\phi_m(\boldsymbol{x})$ is given by:
\begin{equation}
\phi_m(\boldsymbol{x}) = \sigma(\boldsymbol{w}_m \cdot \boldsymbol{x} + b_m).
\end{equation}
with $\boldsymbol{x}\in\Omega$.

In the above equation, $\boldsymbol{w}_m$ (weight vector) and $b_m$ (bias term) are inner parameters that are randomly generated first and then {\it fixed}  throughout the computation process--this makes the problem {\it linear}; $\sigma$ denotes a nonlinear activation function, which provides the necessary nonlinear component for solving the target equation.

In RFM, the loss function is evaluated at collocation points, mirroring the approach of Physics-Informed Neural Networks (PINNs) \cite{raissi2019physics}. It is formulated as a least-squares measure of the equation residuals, augmented with penalty terms to enforce boundary conditions,
\begin{equation}    
Loss = \sum_{\boldsymbol{x}_{i} \in C_{I}}\sum_{k=1}^{K_I}\lambda_{Ii}^{k}\|\mathcal{L}^{k}\boldsymbol{u}(\boldsymbol{x}_{i})-\boldsymbol{f}^{k}(\boldsymbol{x}_{i})\|_{l^{2}}^{2}+ \sum_{\boldsymbol{x}_{j} \in C_{B}}\sum_{\ell=1}^{K_B}\lambda_{Bj}^{\ell}\|\mathcal{B}^{\ell}\boldsymbol{u}(\boldsymbol{x}_{j})-\boldsymbol{g}^{\ell}(\boldsymbol{x}_{j})\|_{l^{2}}^{2}.    
\label{loss2}
\end{equation}

Here, $C_I$ and $C_B$ represent the sets of interior collocation points and boundary collocation points, respectively; $\mathcal{L}^k$ denotes the $k$-th differential operator in the target equation, and $\boldsymbol{f}^k(\boldsymbol{x}_i)$ is the corresponding right-hand-side term at point $\boldsymbol{x}_i$; $\mathcal{B}^{\ell}$ denotes the $\ell$-th boundary operator, and $\boldsymbol{g}^{\ell}(\boldsymbol{x}_j)$ is the corresponding boundary condition value at point $\boldsymbol{x}_j$; $\lambda_{Ii}^k$ and $\lambda_{Bj}^{\ell}$ are the penalty parameters for interior and boundary terms, respectively; $\|\cdot\|_{l^2}$ denotes the $l^2$-norm.

Despite the identical form of the loss function, the optimization problems constructed by RFM and PINNs are fundamentally different. RFM solves a {\it linear} optimization problem: since its inner parameters (i.e., $\boldsymbol{w}_m$ and $b_m$) are fixed, only the outermost linear parameters (i.e., $v_m$) need to be optimized. From the perspective of classical numerical algorithms, RFM transforms the problem into a linear least-squares problem, which can be further represented in the form of the linear system $A\boldsymbol{v} = \boldsymbol{f}$ (where $A$ is the coefficient matrix, $\boldsymbol{v}$ is the vector of linear parameters to be solved, and $\boldsymbol{f}$ is the right-hand-side vector). Consequently, the adjustment of penalty parameters $\lambda_i$ can be directly guided by the information of matrix $A$; more importantly, for each collocation point and each term in the equation, RFM supports the assignment of distinct penalty parameters.

\begin{myexamp}
    
Taking the one-dimensional Helmholtz equation as an example, the loss function of RFM can be written as:

\begin{equation}
\begin{aligned}
L(u) = & \sum_{i=1}^{N-2}\lambda_i\left((\Delta + \lambda^2) u(x_i) - f(x_i)\right)^2 \\
&+ \lambda_0\left(u(x_0)-U(x_0)\right)^2 + \lambda_{N-1}\left(u(x_{N-1})-U(x_{N-1})\right)^2,
\end{aligned}
\end{equation}
where $\Delta$ denotes the second-order differential operator (Laplacian) in one dimension; $\lambda^2$ is the wave number parameter in the Helmholtz equation; $f(x_i)$ is the source term at the $i$-th interior collocation point $x_i$; $U(x_0)$ and $U(x_{N-1})$ are the boundary values at the two endpoints $x_0$ and $x_{N-1}$, respectively; $\lambda_i$ (for $i=1,2,\dots,N-2$), $\lambda_0$, and $\lambda_{N-1}$ are the penalty parameters corresponding to interior points and boundary points.

In practical code implementation, this optimization problem is solved equivalently by constructing the linear system $A\boldsymbol{v} = \boldsymbol{f}$. This construction is based on enforcing the following equalities at all collocation points:
\begin{equation}    
\begin{cases}    
(\Delta + \lambda^2) u(x_i) = f(x_i), & i=1,\cdots,N-2 \\    
u(x_{i}) = U(x_{i}) & i=0,N-1.
\end{cases}    
\end{equation}

\end{myexamp}

\section{The Quantum Random Feature Method}\label{sec:quantumRFM}

In this section, we present the quantum routine for implementing the random feature method. Consistent with conventional quantum PDE solvers, we begin with specific oracles and output a quantum state $\ket{\mathbf{u}}$ that encodes the numerical solution—this excludes the steps of state preparation and measurement. Additionally, a complexity analysis of the algorithm is provided.

\subsection{Oracles}

In the subsequent section, we need following useful oracles.

\begin{myassump}\label{assump:oracle}
    To get the numerical solution of the equation $Lu = f$, suppose we have access to oracles $U_x$, $U_f$, $U_w$ and $U_b$, satisfying:
    \begin{enumerate}
        \item $U_x$ is the block-encoding of the diagonal matrix whose entries are given by the collocation points $x$, i.e.,
        \begin{equation}
            \bra{0^{a_1}} U_{x} \ket{0^{a_1}} = \diag(x), \quad x = [x_0; \cdots; x_{N-1}], \quad N = 2^n.
        \end{equation}

        \item $U_f$ prepares the quantum state $\ket{f(x)}$, i.e.,
        \begin{equation}
            U_{f} \ket{0^n} = \ket{f(x)} = \frac{1}{\left\| f(x) \right\|} \sum_{i \in [2^n]} f(x_i) \ket{i}.
        \end{equation}
        If the point $x_i$ locates on the boundary, we denote $f(x_i)$ as the boundary conditions.

        \item $U_w$ and $U_b$ are block-encodings of randomly generated real, diagonal matrices, i.e.,
        \begin{equation}
            \bra{0^{a_2}} U_{w} \ket{0^{a_2}} = \diag(w), \quad \bra{0^{a_2}} U_{b} \ket{0^{a_2}} = \diag(b), \quad w, b \in \mathbb{R}^{2^m}.
        \end{equation}        
    \end{enumerate}
\end{myassump}

The assumption of the oracle $U_f$ is a standard practice in relevant contexts. Similarly, it is typically assumed that there exists an oracle $O_x$ such that $O_x\ket{0^n} = \ket{x}$. In the present work, however, we assume access to $U_x$; this choice is motivated by the fact that $U_x$ can be constructed with a circuit depth of $O(n)$ and only $O(1)$ queries to $O_x$ or the controlled-$O_x$ \cite{guseynov2025gate, lu2024amplitude}. Furthermore, in scenarios where $x$ exhibits specific architectural characteristics, $U_x$ admits an explicit construction. For illustration, we briefly outline two key procedures: one for constructing $U_x$ from $O_x$, as detailed in \cite{lu2024amplitude}, and another for constructing $U_x$ tailored to the coordinate operator, as presented in \cite{guseynov2024efficient,guseynov2023depth}; additional details are provided in Appendix \ref{appendix:sec:oracle}.

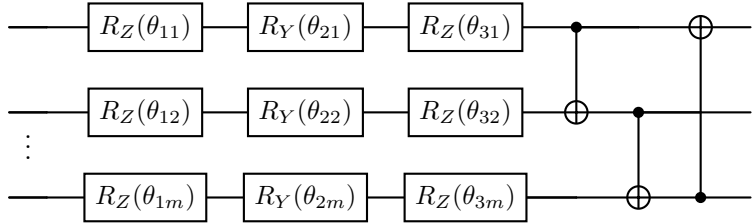
\begin{figure}[!htbp]
    \centering
    \begin{quantikz}
        & \qw & \gate{R_Z(\theta_{11})} & \gate{R_Y(\theta_{21})} & \gate{R_Z(\theta_{31})} & \ctrl{1} &     \qw      &  \targ{}  & \\
        & \qw{\vdots} & \gate{R_Z(\theta_{12})} & \gate{R_Y(\theta_{22})} & \gate{R_Z(\theta_{32})} & \targ{}  & \ctrl{1} &  \qw      & \\
        & \qw & \gate{R_Z(\theta_{1m})} & \gate{R_Y(\theta_{2m})} & \gate{R_Z(\theta_{3m})} &  \qw     & \targ{}  &  \ctrl{-2}& \\
    \end{quantikz}
    \caption{The ansatz for constructing $U_w$ and $U_b$.}
    \label{fig:PQC}
\end{figure}

To obtain oracles $U_{w}$ and $U_{b}$, we first implement the ansatz with randomly sampled parameters. This ansatz $U_{ansatz}$ corresponds to a parameterized quantum circuit (PQC) as defined in Figure \ref{fig:PQC}. Subsequently, $U_w$ and $U_b$ are constructed via the following expression:
\begin{equation}
    P_{\diag} \left( H \otimes I^{\otimes m} \right) \left( \ket{0}\bra{0} \otimes U_{ansatz} + \ket{1}\bra{1} \otimes U_{ansatz}^\dagger \right) \left( H \otimes I^{\otimes m} \right),
\end{equation}
where $P_{\diag}$ projects the original matrix to a diagonal matrix. Adapted from the Hadamard product of block-encodings \cite{guo2024quantum,ivashkov2024qkan}, we have the following lemma stating that $P_{\diag}$ can be implemented using another $m$ ancilla qubits, leading to $a_2=m+1$.
\begin{mylemma}
    Let $U_B$ be an $(\alpha, a, 0)$-block-encoding of $B$, where $B$ is an $m$-qubit operator. Then
    \begin{equation}
        P_{\diag} U_B := \prod_{i=0}^{m-1} \text{CNOT}_{a+m+i}^{a+i} (U_B \otimes I_{m}) \prod_{i=0}^{m-1} \text{CNOT}_{a+m+i}^{a+i},
    \end{equation}
    is an $(\alpha, a+m, 0)$-block-encoding of $\diag(B)$, the diagonal matrix formed by the diagonal entries of $B$. Within the $\text{CNOT}$ gate, the subscript index is used to denote the control qubit.
\end{mylemma}
Therefore, $U_w$ and $U_b$ are $(1,m+1,0)$ block-encodings of random, real and diagonal matrices.

\subsection{Random Feature Basis}
In this subsection, we proceed to construct the analog of the classical random feature basis $\{\phi_j(x)\}_{j=0}^{M-1}$ where $M=2^m$. For the sake of simplicity, we focus on the one-dimensional case with spatial domain $x\in[-1,1]$. Given the collocation points $(x_0,\cdots, x_{N-1})$ as well as the oracles $U_x$, $U_w$, and $U_b$, our objective is to prepare a block-encoding that incorporates the information encapsulated in $\{\phi_j(x_i)\}_{i,j}$.

Following the idea of linear combination of unitaries (LCU) \cite{childs2012hamiltonian}, one can construct $U_{wx+b}$ as:
\begin{equation}
    U_{wx+b} = \left( H \otimes I^{\otimes s} \right) \left(\ket{0}\bra{0} \otimes U_{x} \otimes U_{w} + \ket{1}\bra{1} \otimes I^{\otimes (a_1+n)} \otimes U_{b} \right) \left( H \otimes I^{\otimes s} \right),
\end{equation}
where $s=a_1+n+a_2+m$. Therefore, $U_{wx+b}$ is a $(2,a_1+a_2+1,0)$ block-encoding of $\diag(x) \otimes \diag(w) + I \otimes \diag(b)$. 

\begin{figure}[!htbp]
    \centering
    \begin{quantikz}
        & \qw & \gate{H} & \gate[2]{CR_{\phi_d}} & \qw & \gate[2]{CR_{\phi_{d-1}}} & \qw & \dots & \gate{CR_{\phi_0}} & \gate{H} & \qw \\
        & \qwbundle{a} & \qw & \qw & \gate[2]{U_B} & \qw & \gate[2]{U_B^\dagger} & \dots & \qw & \qw & \qw \\
        & \qwbundle{n} & \qw & \qw & \qw & \qw & \qw & \dots & \qw & \qw & \qw \\
    \end{quantikz}
    \caption{The quantum circuit of $U_\Phi$ using QSVT, where $CR_{\phi}$ is used to implement $e^{i\phi U_\Pi}$ using an ancilla qubit. }
    \label{fig:QSVT}
\end{figure}
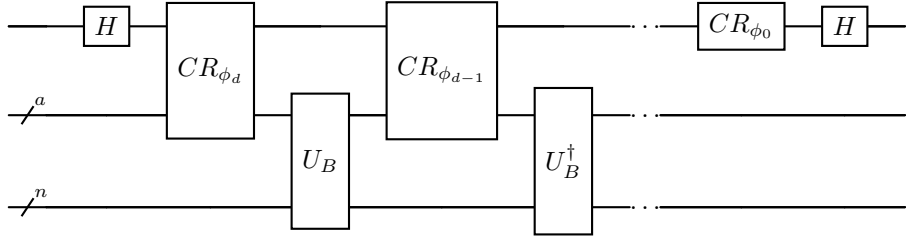

To implement nonlinear activation functions, we employ the widely adopted quantum singular value transformation (QSVT) framework \cite{gilyen2019quantum}. For a given $B \in \mathbb{C}^{N\times N}$ with singular value decomposition $B = W\Sigma V^\dagger$, we define the generalized matrix function as 
\begin{equation}
    f^\diamond(B) := W f(\Sigma) V^\dagger, \quad f^\triangleleft(B) := Wf(\Sigma)W^\dagger, \quad f^\triangleright(B) := Vf(\Sigma)V^\dagger.
\end{equation}
The quantum circuit of QSVT is shown in Figure \ref{fig:QSVT}. Notably, polynomial functions can be applied to the singular values of a block-encoded matrix through QSVT, while smooth functions can be approximated using polynomial representations. For a comprehensive elaboration on QSVT and its applications in quantum algorithms, we refer the reader to Refs. \cite{lin2022lecture, gilyen2019quantum, low2017optimal, low2019hamiltonian, martyn2021grand}.

\begin{mylemma}[Quantum singular value transformation (QSVT)]
    Let $ U_B $ be a $ (1, a, 0) $-general block-encoding of $ B $. Given a polynomial $ P(x) \in \mathbb{R}[x] $ of degree $ d $ satisfying $|P(x)| \leq 1, x\in[-1,1]$ and $P(x)$ has parity $d\mod{2}$, then we can find a sequence of phase factors $ \Phi := (\phi_0, \cdots, \phi_d) \in \mathbb{R}^{d+1} $, so that
    \begin{equation}
        U_{\Phi} := e^{i\phi_0 U_{\Pi}} \prod_{j=1}^{d/2} \left[ U_B^{\dagger} e^{i\phi_{2j-1} U_{\Pi}} U_B e^{i\phi_{2j} U_{\Pi}} \right] = \begin{bmatrix} P^{\triangleright}(B) & * \\ * & * \end{bmatrix}
    \end{equation}
    when $ d $ is even, and
    \begin{equation}
        U_{\Phi} := (-i)^d e^{i\phi_0 U_{\Pi}} \left( U_B e^{i\phi_1 U_{\Pi}} \right) \prod_{j=1}^{(d-1)/2} \left[ U_B^{\dagger} e^{i\phi_{2j} U_{\Pi}} U_B e^{i\phi_{2j+1} U_{\Pi}} \right] = \begin{bmatrix} P^{\diamond}(B) & * \\ * & * \end{bmatrix}
    \end{equation}
    when $ d $ is odd, with $U_\Pi = 2 \ket{0^a} \bra{0^a} - I^{\otimes a} $. The circuit \( U_{\Phi} \) is constructed from \( U_B \), \( U_B^{\dagger} \), an \( a \)-qubit controlled-NOT gate, and \( \mathcal{O}(d) \) single-qubit rotation gates.
\end{mylemma}

When $ B $ is a Hermitian matrix, the singular value transformation coincides with the eigenvalue transformation, i.e., $ P(B) = P^\diamond(B) $. It is important to note that in the present case, $ U_{wx+b} $ serves as the block-encoding of a real diagonal matrix, and thus qualifies as a Hermitian matrix. By applying the QSVT to $ U_{wx+b} $ with the activation function $ \sigma $, we can construct the $ (1, a_1 + a_2 + 2, \varepsilon) $-block-encoding $ U_{\sigma(wx+b)} $ of $ \diag(\phi_j(x_i)) $, where $ \diag(\phi_j(x_i)) $ denotes the values of the random feature basis at the collocation points:
\begin{equation}
    \phi_{j}(x_i) = \sigma(w_j x_i + b_j), \quad i\in[2^n], \quad j\in[2^m].
\end{equation}

\begin{myexamp}\label{examp:tri:poly}
While activation functions like $\tanh$, $\sin$, and $\cos$ are viable for solving PDEs \cite{chen2022bridging}, we focus on the polynomial approximations of trigonometric functions \cite{gilyen2019quantum}, which can be directly implemented via QSVT.
    
Let $ t \in \mathbb{R} \setminus \{0\} $, $ \varepsilon \in \left(0, \frac{1}{e}\right) $, and let $ R = \left\lfloor r\left( \frac{e|t|}{2}, \frac{5}{4}\varepsilon \right) / 2 \right\rfloor $ with $\varepsilon = (t/r)^r$, then the following $ 2R $ and $ 2R + 1 $ degree polynomials satisfy
    \begin{equation}
        \left\| \cos(tx) - J_0(t) + 2\sum_{k=1}^{R} (-1)^k J_{2k}(t) T_{2k}(x) \right\|_{[-1,1]} \leq \varepsilon, \text{ and}    
    \end{equation}
    \begin{equation}
        \left\| \sin(tx) - 2\sum_{k=0}^{R} (-1)^k J_{2k+1}(t) T_{2k+1}(x) \right\|_{[-1,1]} \leq \varepsilon,    
    \end{equation}
    where $ J_m(t): m \in \mathbb{N} $ denote Bessel functions of the first kind and $r(t, \varepsilon) = \Theta\left( t + \log(1/\varepsilon) \right)$.
\end{myexamp}

\subsection{Linear System and Solution}
Given the activation function $\sigma(x)$, we can directly compute its derivative $\sigma^\prime(x)$. This, in turn, allows us to determine the derivative of the random feature basis $\phi_j(x)$ as:
\begin{equation}
     \phi_{j}^\prime(x_i) = w_j \sigma^\prime(w_j x_i + b_j).
\end{equation}
Following the methodology from the previous subsection, we employ QSVT to construct the block-encoding $U_{\sigma^\prime(wx+b)}$ with parameters $(1, a_1+a_2+2, \varepsilon)$. Consequently, the block-encoding for the derivative $\phi^\prime$ is given by:
\begin{equation}
     U_{\phi^\prime} = \left(I^{\otimes 2} \otimes I^{\otimes (a_1+n)} \otimes U_w \right) U_{\sigma^\prime(wx+b)}.
\end{equation}
The same procedure can be generalized to construct block-encodings for higher-order derivatives.

Given the specific form of the linear equation $\mathcal{L}u=f$, LCU can be reutilized to implement the differential operator $\mathcal{L}$, denoted as $U_{\mathcal{L}}$, based on the collocation points distributed within the domain. For collocation points $x_i$ on the boundary, the linear system must strictly comply with the boundary conditions: in the case of Dirichlet conditions, the original form of $\phi_j(x_i)$ is retained directly, while for Neumann or Robin conditions (expressed as $\mathcal{B}u(x_i) = f_i$), the QSVT method is reapplied to construct the corresponding block-encoding $U_\mathcal{B}$. These components are then integrated via control qubits into an extended system, yielding a block-encoding $O_{A}$ that encapsulates the matrix $A$ from the linear system $A\mathbf{v} = \mathbf{f}$. For example, if $x_0$ and $x_{N-1}$ are boundary points, the matrix $A$ takes the following form,
\begin{equation}
    A = [A_{ij}] := \begin{bmatrix}
        \mathcal{B} \phi_0(x_0) & \mathcal{B} \phi_1(x_0) & \cdots & \mathcal{B} \phi_{M-1} (x_0) \\
        \mathcal{L} \phi_0(x_1) & \mathcal{L} \phi_1(x_1) & \cdots & \mathcal{L} \phi_{M-1} (x_1) \\
        \vdots                  & \vdots                  & \ddots                  & \vdots      \\
        \mathcal{L} \phi_0(x_{N-2}) & \mathcal{L} \phi_1(x_{N-2}) & \cdots & \mathcal{L} \phi_{M-1} (x_{N-2}) \\
        \mathcal{B} \phi_0(x_{N-1}) & \mathcal{N} \phi_1(x_{N-1}) & \cdots & \mathcal{B} \phi_{M-1} (x_{N-1}) \\
    \end{bmatrix}.
\end{equation}

It is noteworthy that $O_{A}$ corresponds to the block-encoding of a diagonal matrix with dimension $NM \times NM$. In fact, this $O_{A}$ can be interpreted as an oracle for the matrix 
 $A = [A_{ij}] \in \mathbb{R}^{N \times M}$. The block-encoding of $A$ can be derived using the following lemma:
\begin{mylemma}\label{lemma:oracle2be}
    Given an oracle $O_A$ such that
    \begin{equation}
        O_A \ket{0^l} \ket{i} \ket{j} = A_{ij} \ket{0^l} \ket{i} \ket{j} + \ket{\perp}, \quad i \in [2^n], \quad j \in [2^m],
    \end{equation}
    then 
    \begin{equation}
        U_A = (I^{\otimes l} \otimes I^{\otimes n} \otimes H^{\otimes m}) O_A (I^{\otimes l} \otimes H^{\otimes n} \otimes I^{\otimes m}),
    \end{equation}
    block encodes $A/\sqrt{2^{m+n}}$.
\end{mylemma}
Therefore, we can obtain $U_A$ which is a $(\sqrt{MN}, l^\prime, \varepsilon)$-block-encoding of $A$, where the number of ancilla qubits $l^\prime$ depends on the linear differential operator $\mathcal{L}$, the boudary condition $\mathcal{B}$ and the construction of oracle $U_x$. 

\begin{figure}[!htbp]
    \centering
    \begin{quantikz}
        & \qw & \gate[3]{U_{\tilde{A}}} & \qw \\
        & \qwbundle{l^\prime} & \qw & \qw \\
        & \qwbundle{m} &  & \qw \\
    \end{quantikz}
    \quad := \quad
    \begin{quantikz}
        & \qw & \gate{X} & \ctrl[open]{1} & \ctrl{1} & \qw \\
        & \qwbundle{l^\prime} & \qw & \gate[2]{U_A} & \gate[2]{U_{A}^{\dagger}} & \qw \\
        & \qwbundle{m} & \qw &  &  & \qw \\
    \end{quantikz}
    \\
    \begin{quantikz}
        & \qw & \gate[2]{U_{\tilde{f}}} & \qw \\
        & \qwbundle{m} &  & \qw \\
    \end{quantikz}
    \quad := \quad
    \begin{quantikz}
        & \qw &  & \ctrl[open]{1} &  \\
        & \qwbundle{m} &  &  \gate{U_f}  &  \\
    \end{quantikz}
    
    \caption{The quantum circuit of $U_{\tilde{A}}$ and $U_{\tilde{f}}$.}
    \label{fig:QLSA}
\end{figure}
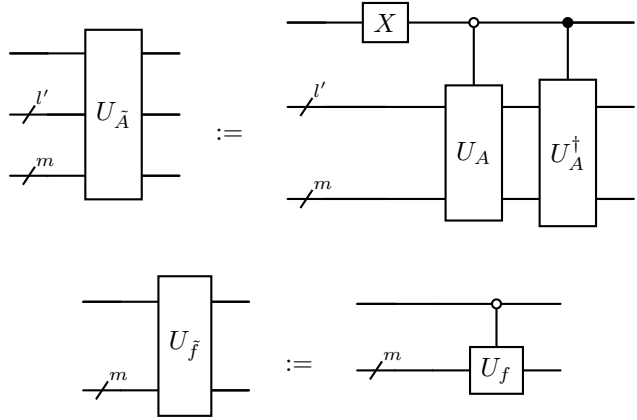

Recalling the idea of the RFM, we aim to find the coefficients of the numerical solution with respect to the random feature basis, namely, to find the quantum state $\ket{\mathbf{v}}$ with $v = [v_j] \in \mathbb{R}^{M}$ satisfying
\begin{equation}
    A \mathbf{v} = \mathbf{f} \quad \Leftrightarrow \quad \begin{bmatrix}
        \mathbf{0} & A \\ A^\dagger & \mathbf{0}
    \end{bmatrix}
    \begin{bmatrix}
        \mathbf{0} \\ \mathbf{v}
    \end{bmatrix}
    =
    \begin{bmatrix}
        \mathbf{f} \\ \mathbf{0}
    \end{bmatrix} \quad \triangleq \quad \tilde{A} \tilde{\mathbf{v}} = \tilde{\mathbf{f}}.
\end{equation}
The block-encoding $U_{\tilde{A}}$ of $\tilde{A}$ and the preparation oracle $U_{\tilde{f}}$ of $\tilde{\mathbf{f}}$ can be obtained by $U_A$ and $U_f$ immediately as shown in Figure \ref{fig:QLSA}. For simplicity, we assume $ M = N $ herein, which implies $ \tilde{\mathbf{v}} \in \mathbb{R}^{2M} $. In cases where $ M \neq N $, the system can be extended by appending zeros.

Many quantum algorithms have been proposed to solve the Quantum Linear System Problem (QLSP). The pioneering work is the well-known Harrow-Hassidim-Lloyd (HHL) algorithm, which has a gate complexity of $\mathcal{O}(\log(N) s^2 \kappa^2 / \varepsilon)$ \cite{harrow2009quantum}, where $s$ is the sparsity and $\kappa$ is the condition number of the matrix $A$. Subsequent research, including QSVT based methods, has aimed to overcome these limitations, improving the scaling with respect to both precision $\varepsilon$ and condition number $\kappa$ \cite{ambainis2012variable,childs2017quantum,gilyen2019quantum,lin2022lecture}. Notably, approaches leveraging adiabatic quantum computing (AQC) \cite{dranov1998discrete,subacsi2019quantum,costa2022optimal} have achieved an optimal query complexity of $\mathcal{O}(\kappa \log(1/\varepsilon))$, independent of the system size $\log(N)$ and sparsity $s$.

To ensure a coherent presentation, we use QSVT as a representative framework to construct the block-encoding of the inverted matrix, denoted as $U_{\tilde{A}^{-1}}$. The random feature coefficients are then solved by the quantum circuit $U_{\tilde{v}} = U_{\tilde{A}^{-1}} O_{\tilde{f}}$, where we have omitted ancillary qubits for simplicity and focus solely on the data register. Subsequently, the solution to the differential equation is constructed via the operation $U_{u} = (I^{\otimes n} \otimes H^{\otimes m}) U_{\sigma(wx+b)} (H^{\otimes n} \otimes I^{\otimes m}) U_{v}$, which physically implements the function approximation $u(x_i) = \sum_{j} v_j \phi_j(x_i)$ at the collocation points.

In summary, the key steps of the proposed algorithm for obtaining the solution state $\ket{\mathbf{u}}$ are outlined as follows:

\begin{enumerate}    
    \item Presume access to the initial oracles $U_x$ and $U_f$, and prepare the block-encodings corresponding to the randomly generated parameters, denoted by $U_w$ and $U_b$.
    
    \item Construct the feature map: Leverage LCU and QSVT to build a block-encoding $U_{\sigma(wx+b)}$ for the random feature basis functions $\phi_j(x_i) = \sigma(w_j x_i + b_j)$.
    
    \item Formulate the linear system: First, construct the block-encodings $U_\mathcal{L}$ (for the linear differential operator) and $U_{\mathcal{B}}$ (for the boundary conditions). Subsequently, combine $U_\mathcal{L}$ and $U_{\mathcal{B}}$ to form $O_{A}$, and further construct a $(\sqrt{MN}, l^\prime, \varepsilon)$-block-encoding $U_A$ for matrix $A$ in the linear system $A \mathbf{v} = \mathbf{f}$. The simulation cost of $U_A$ can then be improved via oblivious amplitude amplification.
    
    \item Solve for unknown coefficients: Embed the aforementioned linear system into a larger Hermitian system $\tilde{A} \tilde{\mathbf{v}} = \tilde{\mathbf{f}}$, and employ a quantum linear system algorithm (e.g., one based on QSVT) to solve for the coefficients $v_j$, which are encoded in the quantum state $\ket{\mathbf{v}}$.
    
    \item Recover the solution state $\ket{\mathbf{u}}$: Apply the pre-constructed feature map $U_{\sigma(wx+b)}$ to the coefficient state $\ket{\mathbf{v}}$. This resulting state encodes the solution $u(x_i) = \sum_j v_j \phi_j(x_i)$ at the collocation points.
\end{enumerate}  

\begin{figure}[!htbp]
    \centering
    \begin{quantikz}
        & & \gate{H} & \gate[2]{CR_{\phi_d}} &  & & & & & \cdots & \\
        &  &  & & \gate{H}\gategroup[wires=3,steps=5,style={inner sep=2pt}]{$U_{wx+b}$} & \ctrl{2} & \ctrl{1} & \ctrl[open]{1} & \gate{H} & \cdots & \\
        & \qwbundle{m} & & & & & \gate{U_w} & \gate{U_b} & & \cdots &  \\
        & \qwbundle{n} & & & & \gate{U_x} & & & & \cdots & 
    \end{quantikz}
    \caption{The quantum circuits of $U_{\sigma(wx+b)}$ using QSVT.}
    \label{fig:circuit:phi}
\end{figure}
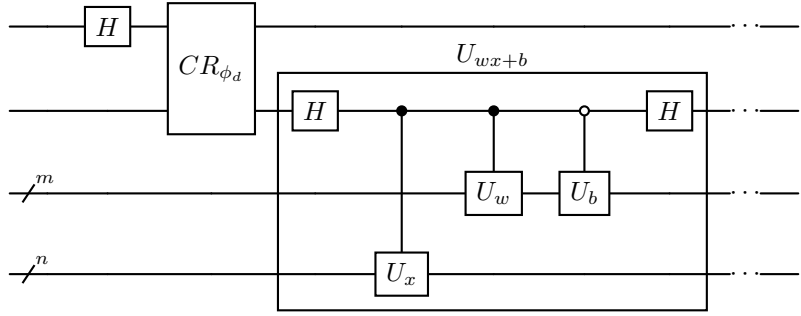

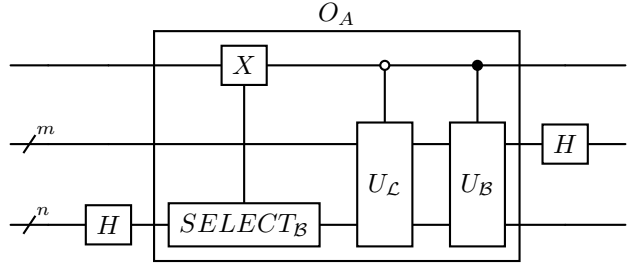
\begin{figure}[!htbp]
    \centering
    \begin{quantikz}
        &  &  & \gate{X}\gategroup[wires=3,steps=3,style={inner sep=2pt}]{$O_{A}$} &  \ctrl[open]{1}  & \ctrl{1} &  &  \\
        & \qwbundle{m} &  &   &  \gate[2]{U_{\mathcal{L}}} & \gate[2]{U_{\mathcal{B}}} &  \gate{H} & \\
        & \qwbundle{n} & \gate{H} &  \gate{SELECT_{\mathcal{B}}} \wire[u][2]{q} &  &  &  &
    \end{quantikz}
    \caption{The quantum circuit for $U_{A}$ is constructed with reference to $U_{\mathcal{L}}$ (the block-encoding of the linear differential operator) and $U_{\mathcal{B}}$ (the block-encoding of boundary conditions). Additionally, the operator $SELECT_{\mathcal{B}}$ specifies the control mechanism when $x_i$ lies on the boundaries.}
    \label{fig:circuit:A}
\end{figure}

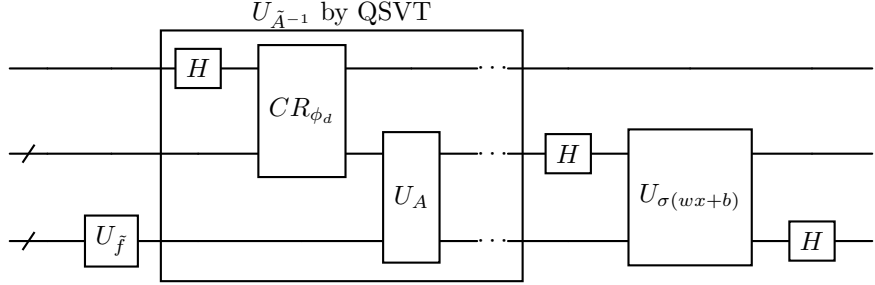
\begin{figure}[!htbp]
    \centering
    \begin{quantikz}
        & &  &  \gate{H}\gategroup[wires=3,steps=4,style={inner sep=2pt}]{$U_{\tilde{A}^{-1}}$ by QSVT} & \gate[2]{CR_{\phi_d}} &  & \cdots &  & &  & \\
        & \qwbundle{} &  &  & & \gate[2]{U_{A}} & \cdots &  \gate{H} & \gate[2]{U_{\sigma(wx+b)}} &  & \\
        & \qwbundle{} & \gate{U_{\tilde{f}}} & &  &  & \cdots &  &  &  \gate{H} & 
    \end{quantikz}
    \caption{The quantum circuit for solving the linear system $\tilde{A} \tilde{v} = \tilde{f}$ and recovering the solution state $\ket{\mathbf{u}}$.}
    \label{fig:circuit:u}
\end{figure}

A diagrammatic illustration of this procedure is provided in Figure \ref{fig:circuit:phi}, Figure \ref{fig:circuit:A}, and Figure \ref{fig:circuit:u}.

\subsection{Analysis}

The following theorem establishes the asymptotic complexity of the proposed quantum random feature method.
\begin{mythm}
    Given the oracles \( U_x \), \( U_f \), \( U_w \), and \( U_b \) as defined in Assumption \ref{assump:oracle}, the quantum random feature method requires \( O(m+n) \) ancilla qubits. The total gate complexity for preparing the solution state \( \ket{\mathbf{u}} \) of the differential equation \( \mathcal{L}u = f \), including queries to these oracles and the number of additional single- and two-qubit gates, is
\begin{equation}
O\left( \sqrt{MN} \cdot \poly(mn) \cdot \log \frac{1}{\varepsilon} \right),
\end{equation}
where \( N = 2^n \) is the number of collocation points, \( M = 2^m \) is the number of random feature bases, and \( \varepsilon \) is the target precision.
\end{mythm}

It is noteworthy that in the classical RFM, the construction of the matrix $A$ and the solution of the resulting linear system each require $\Omega(MN)$ operations. Consequently, the proposed quantum algorithm achieves a  {\it quadratic} speedup.

Compared to other quantum methods built on numerical schemes such as Schr\"odingerisation \cite{jin2022quantum}, this approach employs neural network-like basis functions in place of traditional tensor-product based basis functions. A key implication is that the number of terms required does not necessarily scale as $N^d$, where $N$ denotes the number of unknown parameters per dimension and $d$ represents dimensionality. Instead, this number depends solely on the complexity of the solution. While quantum computing enables the encoding of $N^d$ parameters into $d\log N$ qubits, such a reduction remains significant, particularly for near-term quantum devices with limited resources. For conventional methods, the difference matrix is typically sparse, yet its condition number is dependent on $N$ and can become extremely large. In contrast, when using QRFM, the condition number of matrix $A$ does not explicitly rely on $N$. Although matrix $A$ is generated using random parameters $w$ and $b$, once $w$ and $b$ are fixed, the penalty parameter $\lambda$ can be adjusted to modify matrix $A$.

\begin{proof}

To start with, the resource requirements for the oracles defined in Assumption \ref{assump:oracle} are as follows (see Appendix \ref{appendix:sec:oracle} and Figure \ref{fig:PQC}). The oracle \( U_x \) requires \( O(n) \) ancilla qubits, \( O(1) \) queries to \( O_x \), and \( O(n^2) \) additional single- and 2-qubit gates. The oracles \( O_w \) and \( O_b \) require \( O(m) \) ancilla qubits and \( O(m) \) single- and 2-qubit gates.

The dominant computational cost in constructing the random feature basis and the linear system arises from the QSVT procedure, which scales linearly with both the cost of constructing block-encodings for the input matrices and the degree of the polynomials involved. According to \cite[Corollary 23]{gilyen2019quantum}, a piecewise smooth function can be approximated by a polynomial of degree \( O(\log(1/\varepsilon)) \), where \( \varepsilon \) is the approximation error. For example, the trigonometric activation functions in Example \ref{examp:tri:poly} are approximated by polynomials of degree \( O(\log(1/\varepsilon)) \).

The total complexity of constructing the unitaries \( U_\mathcal{L} \) and \( U_{\mathcal{B}} \) depends on the structure of the differential operator \( \mathcal{L} \) and the boundary condition operator \( \mathcal{B} \). Assuming both \( \mathcal{L} \) and \( \mathcal{B} \) contain \( O(1) \) terms with the highest derivative order being \( O(1) \), the unitary \( O_A \) requires:
- \( O(m+n) \) ancilla qubits,
- \( O(\poly(mn)\log(1/\varepsilon)) \) queries to the oracles \( U_x \), \( U_w \), and \( U_b \), and
- \( O(\poly(mn)) \) additional single- and 2-qubit gates.

By Lemma \ref{lemma:oracle2be}, this yields an \( (\sqrt{MN}, l, \varepsilon) \)-block-encoding of \( A \). Applying oblivious amplitude amplification \cite{gilyen2019quantum} then produces a \( (1, l+1, \sqrt{MN}\varepsilon) \)-block-encoding, which we denote as \( U_A \) for simplicity. This step uses a single ancilla qubit, \( O(\sqrt{MN}) \) calls to \( U_A \) and \( U_A^\dagger \), and \( O(\sqrt{MN}) \) additional single- and 2-qubit gates.

To solve the QLSP such that the normalized output state is \( \varepsilon \)-close to the normalized solution state \( \ket{\tilde{v}} \), the block-encoding of \( U_{\tilde{A}^{-1}} \) must be prepared with precision \( O(\varepsilon \xi / \kappa) \), where \( \kappa \) is the condition number and \( \xi := \| \tilde{A}^{-1} \ket{\tilde{f}} \| \). The success probability of this procedure is \( \Omega(\xi^2 / \kappa^2) \), which can be boosted to \( \Omega(1) \) with \( O(\kappa / \xi) \) rounds of amplitude amplification \cite{lin2022lecture}. The overall number of queries to \( U_A \) and its inverse is \( O(\kappa^2 \log(\kappa/\varepsilon)) \), leading to a total oracle query complexity of
\begin{equation}
O\left( \sqrt{MN} \cdot \poly(mn) \cdot \log \frac{1}{\varepsilon} \right)
\end{equation}
for the oracles \( U_x \), \( U_w \), \( U_b \), and \( U_f \).
\end{proof}

\section{Numerical Experiments}\label{sec:experiment}

This section provides a numerical example to demonstrate the proposed quantum random feature method, where the block-encodings \( U_{\mathcal{L}} \) and \( U_{\mathcal{B}} \) are constructed from the explicit form of the differential equation and implemented in \textit{PennyLane}. Although the algorithm's complexity scales favorably as \( O(\log(1/\varepsilon)) \), current quantum hardware lacks the precision required for practical execution, underscoring the need for advancements in quantum error correction and error mitigation techniques. Due to computational constraints, our simulation classically emulates the quantum process: the weights \( \mathbf{w} \) and biases \( \mathbf{b} \) are derived from the matrix representations of \( U_w \) and \( U_b \), and the QSVT procedures are replaced with exact matrix arithmetic.

Consider the one-dimensional Helmholtz equation on the domain $\Omega = [-1,1]$ with Dirichlet boundary conditions:
\begin{equation}
\begin{cases}
\dfrac{\mathrm{d}^2 u(x)}{\mathrm{d} x^2} + k^2 u(x) = f(x), & x \in \Omega, \\
u(-1) = c_1, \quad u(1) = c_2,
\end{cases}
\end{equation}
where the wave number is $k=2$. We select an exact solution of the form
\begin{equation}
    u(x) = \sin\big(3\pi(x + 0.05)\big) \cos\big(2\pi(x + 0.05)\big) + 2,
\end{equation}
from which the boundary values $c_1$, $c_2$, and the forcing function $f(x)$ can be directly computed.

The block-encoding $U_{\mathcal{L}}$ depends on the choice of activation function $\sigma$, for example,
\begin{itemize}
    \item for $\sigma = \sin$ or $\sigma = \cos$, the operator $\mathcal{L}$ acting on the feature map yields
    \begin{equation}
    \mathcal{L}u = (k^2 - w^2) \sigma(wx+b),
    \end{equation}
    which implies the block-encoding can be constructed as $U_{\mathcal{L}} = U_{k^2 - w^2} U_{\sigma(wx+b)}$. Here, $U_{k^2-w^2}$ is itself built from $U_w$ via QSVT.
    
    \item for $\sigma = \tanh$, the corresponding expression becomes
    \begin{equation}
    \mathcal{L}u = 2w^2 \sigma^3(wx+b) + (k^2 - 2w^2) \sigma(wx+b).
    \end{equation}
\end{itemize}

In numerous prior studies on quantum neural networks (QNNs), a notable issue arises: as the depth of the QNN and the number of its parameters increase, the gradients of the model often tend to vanish, which is known as the ``barren plateau'' \cite{cerezo2021cost, grant2019initialization}. By contrast, in the RFM, parameter training is unnecessary. Furthermore, weight enhancement can be readily achieved by selecting $\sigma(\alpha(wx+b))$ as the basis, where $\alpha$ serves as a parameter to promote gradient improvement.

We employ $N = 2^8$ collocation points uniformly distributed across the domain, with $x_0 = -1$ and $x_{N-1} = 1$. The solution is approximated using a set of basis functions, where the number of basis functions is $M = 2^m$ for $m = 3, 4, 5, 6, 7$. In the quantum circuit implementation, the block-encoding of the boundary operator, $U_{\mathcal{B}}$, is applied to the quantum states $\ket{0^n}$ and $\ket{1^n}$, while the block-encoding of the differential operator, $U_{\mathcal{L}}$, is applied to the remaining quantum states.

\begin{figure}[!htbp]
  \centering
  \begin{subfigure}{0.45\textwidth}
    \centering
    \includegraphics[width=\textwidth]{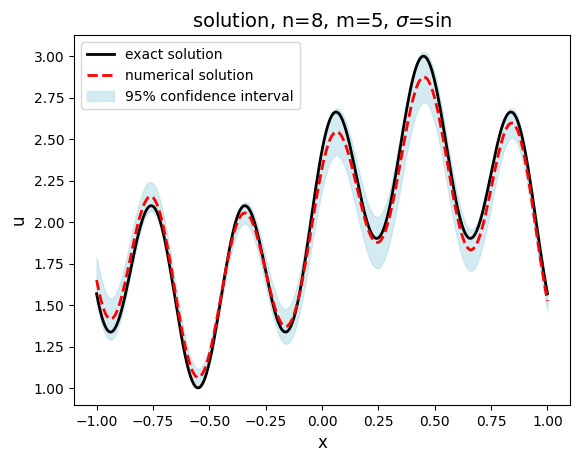}
  \end{subfigure}
  \hspace{1cm}
  \begin{subfigure}{0.45\textwidth}
    \centering
    \includegraphics[width=\textwidth]{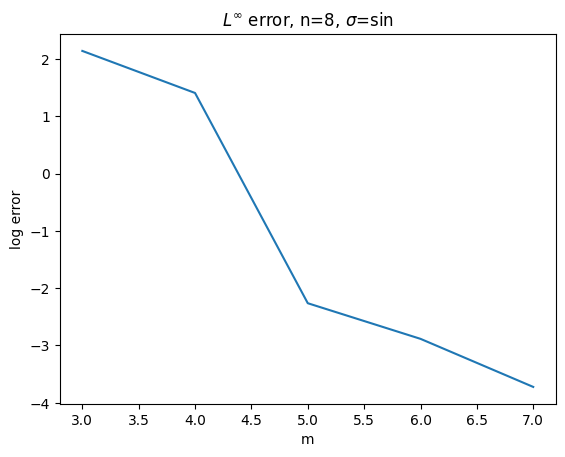}
  \end{subfigure}
  \hspace{1cm}
  \begin{subfigure}{0.45\textwidth}
    \centering
    \includegraphics[width=\textwidth]{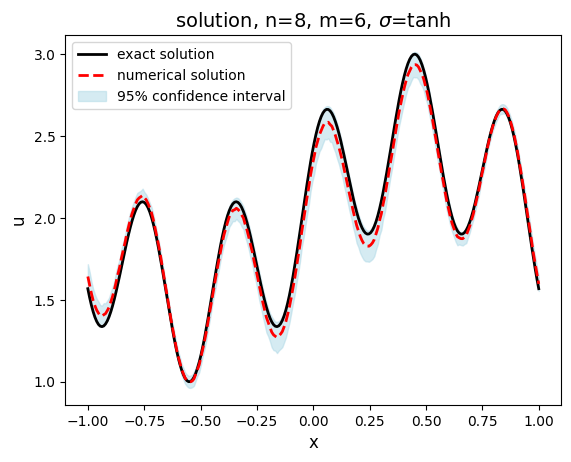}
  \end{subfigure}
  \hspace{1cm}
  \begin{subfigure}{0.45\textwidth}
    \centering
    \includegraphics[width=\textwidth]{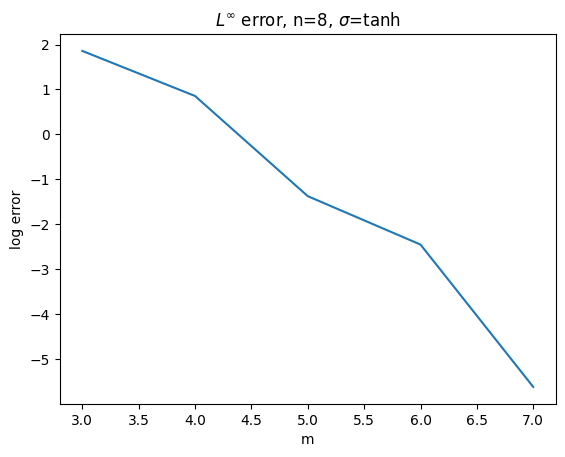}
  \end{subfigure}
  \caption{Numerical solutions and $\log$ $L^\infty$ errors of the Helmholz equation, using $\sin$ and $\tanh$ as the activation function. $n$ is defined as the number of data qubits (with $N=2^n$ representing the number of mesh grid points), and $m$ is defined as the number of random feature bases.}
  \label{fig:helmholz}
\end{figure}

Numerical results are presented in Figure~\ref{fig:helmholz}. For each value of $m$, we perform 100 independent trials to compute the confidence intervals for the numerical solution. As shown in the figure, the method provides an accurate approximation of the exact solution. Furthermore, the numerical solution converges as $m$ increases, demonstrating consistent performance for both $\sin$ and $\tanh$ activation functions.

\section{Dual formulation: quantum kernel regression}
\label{sec:kernel}

Here, we highlight a dual interpretation of our method through the lens of kernel methods, a cornerstone of nonlinear statistical learning \cite{scholkopf2002learning,paine2023quantum,rahimi2007random,liu2021random}. Consider two samples \( x, x^\prime \in \mathcal{X} \), and a nonlinear feature map \( \phi : \mathcal{X} \mapsto \mathcal{H} \) into a reproducing kernel Hilbert space (RKHS) \( \mathcal{H} \). The inner product in \( \mathcal{H} \) between the mapped features \( \phi(x) \) and \( \phi(x^\prime) \) defines a kernel function:
\begin{equation}
k(x, x^\prime) = \langle \phi(x), \phi(x^\prime) \rangle_{\mathcal{H}}.
\end{equation}

In practice, the kernel function \( k \) is used directly to compute the inner product via the \textit{kernel trick}, which avoids the need for an explicit feature mapping \( \phi \). However, for a dataset of \( N \) samples in \( \mathcal{X} \), kernel ridge regression (KRR) becomes computationally prohibitive, requiring \( O(N^3) \) time for training and \( O(N^2) \) space for the kernel matrix. These costs render standard KRR infeasible for large \( N \).

To overcome this, Random Fourier Features (RFF) provides an efficient kernel approximation. Based on Bochner's Theorem \cite{bochner2005harmonic}, which states that shift-invariant kernels can be represented as the Fourier transform of a probability distribution, the kernel is approximated by:
\begin{equation}
k(x, x') \approx \tilde{k}_p(x, x') := \varphi_p(x)^T \varphi_p(x'),
\end{equation}
where the expectation \( \mathbb{E}_{\omega \sim p}[\varphi_p(x)^T \varphi_p(x')] \) converges to the true kernel. The explicit feature mapping is:
\begin{equation}
\varphi_p(x) := \frac{1}{\sqrt{M}} \left[ \exp(-\mathrm{i} \omega_0 x), \cdots, \exp(-\mathrm{i} \omega_{M-1} x) \right]^T,
\end{equation}
with frequencies \( \{\omega_j\}_{j=0}^{M-1} \) sampled independently from the distribution \( p(\cdot) \), independent of the training data.

Consequently, the original \( N \times N \) kernel matrix \( K = [k(x_i, x_j)]_{N \times N} \) is approximated by \( \widetilde{K}_p = Z_p Z_p^T \), where \( Z_p = [\varphi_p(x_0), \cdots, \varphi_p(x_{N-1})]^T \in \mathbb{R}^{N \times M} \) is the matrix of random Fourier features for the training data. For a detailed survey, see \cite{liu2021random}.

The kernel regression formulation is equivalent to the proposed QRFM. In fact, the classical method is a special case where the quantum feature mapping takes the explicit form \( \phi_{j}(x) = \exp(-\mathrm{i} \omega_j x) / \sqrt{M} \). Following a procedure analogous to the classical random feature method, the solution \( u \) can be derived as:
\begin{equation}
u(x) = \sum_{l=0}^{N-1} \alpha_{l} k(x, x_{l}) = \sum_{l=0}^{N-1} \alpha_{l} \sum_{j=0}^{M-1} \phi_j(x) \phi_{j}(x_l) = \sum_{j=0}^{M-1} \underbrace{\left(\sum_{l=0}^{N-1} \alpha_{l} \phi_{j}(x_l) \right)}_{\beta_j} \phi_{j}(x).
\end{equation}
This reveals that the solution can be interpreted as a linear model \( u(x) = \sum_{j=0}^{M-1} \beta_j \phi_j(x) \) in the feature space defined by \( \{\phi_j\} \).

Both methods require optimizing a dual quadratic loss function, which necessitates solving quantum linear algebra problems. The primal formulation (the proposed method) solves for the state \( \ket{\mathbf{v}} \), while the dual formulation (kernel regression) solves for \( \ket{\boldsymbol{\alpha}} \). These solutions are related by the transformation:
\begin{equation}
v_j = \sum_{l=0}^{N-1} \alpha_l \phi_j(x_l), \quad \text{or equivalently}, \quad \mathbf{v} = Z_p^T \boldsymbol{\alpha}.
\end{equation}

Once the state \( \ket{\boldsymbol{\alpha}} \) is prepared, the block-encoding \( U_K \) of the kernel matrix is applied to produce the solution state \( \ket{\mathbf{u}} \). In cases where \( U_K \) can be constructed efficiently, such as for polynomially or exponentially decaying kernels, the computational complexity can be reduced to \( O(\poly\log(N)) \) \cite{nguyen2022block}. A more precise analysis of which formulation is more advantageous in specific problem settings is deferred to future work on particular PDEs.

\section{Discussion}\label{sec:discussion}

This paper introduces a quantum random feature method (QRFM) to address the high computational cost of solving (high-dimensional) PDEs. The method generates random feature functions and solves the resulting system on a quantum computer. The QRFM significantly outperforms its classical counterpart, achieving at least a quadratic reduction in computational complexity. By leveraging amplitude encoding to manipulate solution states directly, the approach aligns with the intrinsic advantages of quantum computing and facilitates integration with other quantum algorithms.

Several promising research directions remain. First, designing quantum circuits for multi-scale or localized random feature bases (see Appendix \ref{appendix:sec:local}) could better capture localized solution structures. Second, quantum preprocessing techniques, such as those based on quantum preconditioners \cite{jin-precondition}, could mitigate bottlenecks from ill-conditioned linear systems. Third, optimizing the ansatz for quantum weight generation may improve computational accuracy. Extending the method to nonlinear PDEs and validating it on real-world problems are also critical next steps. Finally, there is room to explore novel quantum feature mappings \( \phi(x) \) in the random feature method. In addition, through the dual formalism, new quantum feature mappings could yield kernels that are more efficient to compute than classical alternatives. These extensions may have the potential to advance quantum numerical methods for high-dimensional PDEs.

\section*{Acknowledgement}
SJ, NL and LZ are supported by NSFC grant No. 12341104, the Shanghai Jiao Tong University 2030 Initiative and the Fundamental Research Funds for the Central Universities. SJ is also supported by NSFC grant No. 92270001. NL also acknowledges funding from the Science and Technology Commission of Shanghai Municipality (STCSM) grant no. 24LZ1401200 (21JC1402900) and NSFC grant No. 12471411, and the Shanghai Science and Technology Innovation Action Plan (24LZ1401200).  LZ is also partially supported by the NSFC grant No. 12271360.

\bibliographystyle{ieeetr}
\bibliography{ref}

\appendix
\section{Local Random Feature Function and Partition of Unity}\label{appendix:sec:local}

The random feature functions in Section \ref{sec:RFM} are defined globally; however, in certain scenarios, the solutions to differential equations exhibit small-scale local variations. Consequently, RFM may also involve constructing local random feature functions across multiple local regions, which are subsequently combined via the partition of unity technique.

Specifically, first, we select the centers $\{\boldsymbol{x}_n\}_{n=1}^{M_p} \subset \Omega$ of the partition of unity functions. For these $ M_p $ local regions, we construct affine transformations:
\begin{equation}
    \tilde{\boldsymbol{x}} = \frac{1}{\boldsymbol{r}_{n}}(\boldsymbol{x} - \boldsymbol{x}_{n}), \quad n=1, \cdots, M_p.
\end{equation}
This affine transformation maps the small local region $[x_{n1}-r_{n1}, x_{n1}+r_{n1}] \times \cdots \times [x_{nd}-r_{nd}, x_{nd}+r_{nd}]$ to the unified interval $[-1,1]^{d}$, facilitating the fitting of local features. The construction of the partition of unity functions relies on this affine transformation, and they are often chosen as:
\begin{equation}
    \psi_{n}^{a}(x) = \mathbb{I}_{-1 \leq \tilde{x} < 1},
    \label{psi1}
\end{equation}
or
\begin{equation}
    \psi_{n}^{b}(x) = \mathbb{I}_{\left[-\frac{5}{4}, -\frac{3}{4}\right]}(\tilde{x}) \frac{1 + \sin(2 \pi \tilde{x})}{2} + \mathbb{I}_{\left[-\frac{3}{4}, \frac{3}{4}\right]}(\tilde{x}) + \mathbb{I}_{\left[\frac{3}{4}, \frac{5}{4}\right]}(\tilde{x}) \frac{1 - \sin(2 \pi \tilde{x})}{2}.
\end{equation}

Subsequently, in each local region, we define $ J_n $ random feature functions:
\begin{equation}\label{eqn:basis0}
    \phi_{nj}(\boldsymbol{x}) = \sigma(\boldsymbol{k}_{nj} \cdot \tilde{\boldsymbol{x}} + b_{nj}), \quad j=1, \cdots, J_n.
\end{equation}
Then, the final numerical solution is obtained by combining the local random feature functions through the partition of unity functions:
\begin{equation}
    u_M(\boldsymbol{x}) = \sum_{n=1}^{M_p} \psi_n (\boldsymbol{x}) \sum_{j=1}^{J_n} u_{nj} \phi_{nj} (\boldsymbol{x}).
    \label{representation2}
\end{equation}

It is important to note that if the partition of unity functions $\psi^{a}$ lack the continuity inherent to the governing equation, additional continuity constraints must be enforced. For instance, when solving a second-order partial differential equation, the functions $\psi^{a}$ must be augmented with zeroth-order and first-order continuity conditions across the interfaces of adjacent computational subdomains.


\section{Construction of Oracle $U_x$}\label{appendix:sec:oracle}
\subsection{Cyclic Matrix Representation}
Let $ F $ denote the Fourier matrix of size $ \mathbb{C}^{N \times N} $, where $ \omega = e^{\frac{2\pi i}{N}} $ denotes the $ N $-th root of unity. The matrix $ F $ admits the following explicit form:
\begin{equation}
    F = \frac{1}{\sqrt{N}} \begin{pmatrix}
    1 & 1 & 1 & \cdots & 1 \\
    1 & \omega & \omega^2 & \cdots & \omega^{N-1} \\
    \vdots & \vdots & \vdots & \ddots & \vdots \\
    1 & \omega^{N-1} & \omega^{2(N-1)} & \cdots & \omega^{(N-1)^2}
    \end{pmatrix}.
\end{equation}
For a diagonal matrix $ A = \diag(\mathbf{x}) $ with $ \mathbf{x} \in \mathbb{C}^{N} $, a cyclic matrix $ A^\prime $ can be constructed via the Fourier matrix $ F $. Specifically, $ A^\prime $ satisfies:
\begin{equation}
    A^\prime = \sqrt{N} F^{-1} A F,
\end{equation}
and has the cyclic structure:
\begin{equation}
    A^\prime = \begin{pmatrix}
    a_0 & a_1 & a_2 & \cdots & a_{N-1} \\
    a_{N-1} & a_0 & a_1 & \cdots & a_{N-2} \\
    \vdots & \vdots & \vdots & \ddots & \vdots \\
    a_1 & a_2 & a_3 & \cdots & a_0
    \end{pmatrix}.
\end{equation}
Here, the vector $ \mathbf{a} = [a_0; a_1; \cdots; a_{N-1}] \in \mathbb{C}^N $ is defined as the product of the Fourier matrix $ F $ and the vector $ \mathbf{x} $, i.e., $ \mathbf{a} = F \mathbf{x} $.

The cyclic matrix $ A^\prime $ can be expressed as a linear combination of the matrix $ U $, which is given by:
\begin{equation}    
    A^\prime = \sum_{i=0}^{N-1} a_i U^i, \quad U = \begin{pmatrix}    
    0 & 1 & 0 & \cdots & 0 \\    
    0 & 0 & \ddots & \ddots & \vdots \\    
    \vdots & \vdots & \ddots & \ddots & 0 \\    
    0 & 0 & \cdots & 0 & 1 \\    
    1 & 0 & \cdots & 0 & 0    
    \end{pmatrix},
\end{equation}
where $ U $ denotes the cyclic shift matrix. By virtue of the LCU technique \cite{childs2012hamiltonian}, the $(1,n,0)$-block-encoding $ U_{\frac{1}{\sqrt{N}}A^\prime} $ can be defined as follows:
\begin{equation}\label{appendix:eqn:lcu}
    U_{\frac{1}{\sqrt{N}}A^\prime} = \left(H^{\otimes n} \otimes I^{\otimes n}\right) U_{SET} \left(U_{PREP} \otimes I^{\otimes n}\right),
\end{equation}
in which $ H $ represents the Hadamard gate and $ I $ denotes the identity matrix. Specifically:
\begin{equation}    
    U_{SET} = \sum_{i=0}^{N-1} \ket{i}\bra{i} \otimes U^i, \quad U_{PREP} \ket{0^n} = \sum_{i=0}^{N-1} a_i \ket{i}.
\end{equation}

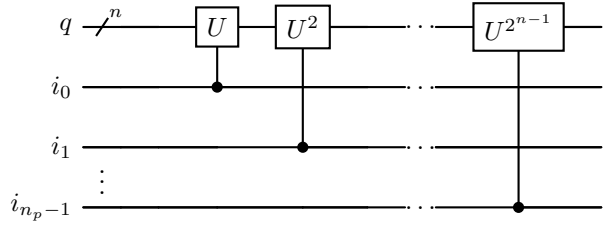
\begin{figure}[!htbp]
    \centering
    \begin{quantikz}[row sep={0.8cm,between origins}]
        \lstick{$q$} & \qwbundle{n} & \qw & \gate{U} & \gate{U^2} & \qw & \ldots & \gate{U^{2^{n-1}}} & \qw \\
        \lstick{$i_0$} & \qw & \qw & \ctrl{-1} & \qw & \qw & \ldots & \qw & \qw \\
        \lstick{$i_1$} & \qw{\vdots} & \qw & \qw & \ctrl{-2} & \qw & \ldots & \qw & \qw \\
        \lstick{$i_{n_p-1}$} & \qw & \qw & \qw & \qw & \qw & \ldots & \ctrl{-3} & \qw \\
    \end{quantikz} 
    \caption{Quantum circuit for $U_{SET}$.}
    \label{fig:Uset}
\end{figure}

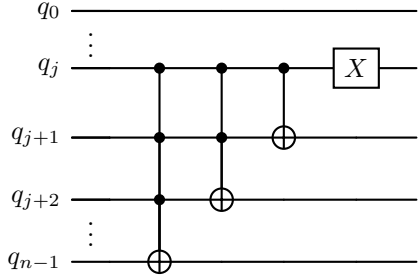
\begin{figure}[!htbp]
    \centering
    \begin{quantikz}
        \lstick{$q_0$} & \qw{\vdots} &&&&& \\
        \lstick{$q_{j}$} & \qw & \ctrl{3} & \ctrl{2} & \ctrl{1} & \gate{X} & \\
        \lstick{$q_{j+1}$} & \qw &  \ctrl{2} & \ctrl{1} & \targ{} &  & \\
        \lstick{$q_{j+2}$} & \qw{\vdots} & \ctrl{1} & \targ{} &  &  & \\
        \lstick{$q_{n-1}$} & &  \targ{} & & & & 
    \end{quantikz}
    \caption{The quantum circuit of $U^{2^{j}}$.}
    \label{fig:Uj}
\end{figure}

As stated in \cite[Lemma 3]{hu2024circuit}, the unitary $ U_{SET} $ can be realized by applying the controlled operator $ c\text{-}U^{2^j} $ to the second register, with the control signal derived from the $ j $-th qubit of the first register. Furthermore, the unitary $ U^{2^j} $ can be implemented via the following expression:
\begin{equation}    
U^{2^{j}} = X_{j} \prod_{k=1}^{n-j-1} \text{CNOT}_{j,\cdots,n-k-1}^{n-k},
\end{equation}
where $ X_j $ denotes the Pauli-X gate acting on the $ j $-th qubit, and $ \text{CNOT}_{j,\cdots,n-k-1}^{n-k} $ represents the multi-controlled-NOT (MCNOT) gate with the specified control and target qubits. The corresponding quantum circuits for $ U_{SET} $ and $ U^{2^j} $ are illustrated in Figure \ref{fig:Uset} and Figure \ref{fig:Uj}, respectively.

Given the oracle $O_{\mathbf{x}}$, it is evident that $U_{PREP} = F O_{\mathbf{x}}$, where $F$ is realized via the quantum Fourier transform. Finally, the oracle $U_{x}$ is defined as follows:
\begin{equation}
    U_{x} = (I^{\otimes n} \otimes F) U_{\frac{1}{\sqrt{N}}A^\prime} (I^{\otimes n} \otimes F^{-1}).
\end{equation}

\subsection{Block-encoding of the Coordinate Operators}

Assume the collocation points form a uniform grid on the interval \([-1,1]\), given by
\begin{equation}
x_i = -1 + \frac{2i}{2^n - 1}, \quad i = 0, \cdots, 2^n-1.
\end{equation}
The corresponding diagonal matrix \(\hat{x} = \diag(x_i)\) can be expressed as
\begin{equation}
\hat{x} = \sum_{i=0}^{2^n-1} \left( -1 + \frac{2i}{2^n-1} \right) \ket{i}\bra{i}.
\end{equation}

Using the binary representation \(\ket{i} = \ket{i_{n-1}\cdots i_0}\), we derive a tensor product form. First, note that \(i = \sum_{k=0}^{n-1} i_k 2^k\). Substituting this into the expression for \(\hat{x}\) yields:
\begin{equation}
\begin{aligned}
\hat{x} &= -I^{\otimes n} + \frac{2}{2^n-1} \sum_{i_{n-1},\cdots,i_0=0}^{1} \sum_{k=0}^{n-1} i_k 2^k \ket{i_{n-1}}\bra{i_{n-1}} \otimes \cdots \otimes \ket{i_0}\bra{i_0} \\
&= - I^{\otimes n} + \frac{2}{2^n-1} \sum_{k=0}^{n-1} 2^k \sum_{i_{n-1},\cdots,i_0=0}^{1} i_k \ket{i_{n-1}}\bra{i_{n-1}} \otimes \cdots \otimes \ket{i_0}\bra{i_0}.
\end{aligned}
\end{equation}

The inner sum acts as the identity on all qubits except the \(k\)-th, where it projects onto \(\ket{1}\bra{1}\). Thus,
\begin{equation}
\sum_{i_{n-1},\cdots,i_0=0}^{1} i_k \ket{i_{n-1}}\bra{i_{n-1}} \otimes \cdots \otimes \ket{i_0}\bra{i_0} = I^{\otimes (n-1-k)} \otimes \ket{1}\bra{1} \otimes I^{\otimes k}.
\end{equation}

Substituting back, we have:
\begin{equation}
\hat{x} = -I^{\otimes n} + \frac{2}{2^n-1} \sum_{k=0}^{n-1} 2^k \left( I^{\otimes (n-1-k)} \otimes \ket{1}\bra{1} \otimes I^{\otimes k} \right).
\end{equation}

Noting that \(I^{\otimes n} = \sum_{k=0}^{n-1} \frac{2^k}{2^n-1} I^{\otimes n}\) and \(I - 2\ket{1}\bra{1} = Z\), we can rewrite the expression as:
\begin{equation}
\begin{aligned}
\hat{x} &= -\sum_{k=0}^{n-1} \frac{2^k}{2^n-1} \left[ I^{\otimes n} - 2\left( I^{\otimes (n-1-k)} \otimes \ket{1}\bra{1} \otimes I^{\otimes k} \right) \right] \\
&= -\sum_{k=0}^{n-1} \frac{2^k}{2^n-1} \left( I^{\otimes (n-1-k)} \otimes Z \otimes I^{\otimes k} \right) \\
&= -\sum_{k=0}^{n-1} \frac{2^k}{2^n-1} Z_k,
\end{aligned}
\end{equation}
where \(Z_k \equiv I^{\otimes (n-1-k)} \otimes Z \otimes I^{\otimes k}\).

Then LCU allows us to construct a \((1, \log_2 n, 0)\)-block-encoding \(U_x\) of \(\hat{x}\) as follows:
\begin{equation}
\begin{aligned}
U_{x} &= - \left(U_{PREP}^\dagger \otimes I^{\otimes n}\right) U_{SET} \left(U_{PREP} \otimes I^{\otimes n}\right), \\
U_{SET} &= \sum_{j=0}^{n-1} \ket{j}\bra{j} \otimes Z_j, \quad U_{PREP} \ket{0^{\log_2 n}} = \sum_{j=0}^{n-1} \sqrt{\frac{2^j}{2^{n}-1}} \ket{j}.
\end{aligned}
\end{equation}

\end{document}